 \documentclass[prl,twocolumn,showpacs]{revtex4}
 \usepackage{graphicx}
 \usepackage{dcolumn}
 \usepackage{bm}
 
\begin{document}

 \title{Phase-field Crystals with Elastic Interactions}
 \author{Peter Stefanovic$^1$, Mikko Haataja$^2$, Nikolas Provatas$^1$}
 \affiliation{$^1$Department of Materials Science and Engineering and Brockhouse 
 Institute for Materials Research, McMaster University, $^2$Department of Mechanical and Aerospace
 Engineering and the Princeton Institute for the Science and
 Technology of Materials (PRISM), Princeton University}
 \date{\today}
 
 \begin{abstract}
 We report on a novel extension of the recent phase-field crystal
 (PFC) method introduced in [Elder {\it et al.}, Phys. Rev. Lett.,
 {\bf 88}, 245701:1-4 (2002)], which incorporates elastic
 interactions as well as crystal plasticity and diffusive dynamics.
 In our model, elastic interactions are mediated through wave modes
 that propagate on time scales many orders of magnitude slower than
 atomic vibrations but still much faster than diffusive times
 scales. This allows us to preserve the quintessential advantage of
 the PFC model: the ability to simulate atomic-scale interactions
 and dynamics on time scales many orders of magnitude longer than
 characteristic vibrational time scales. We demonstrate the two
 different modes of propagation in our model and show that
 simulations of grain growth and elasto-plastic deformation are
 consistent with the microstructural properties of nanocrystals.
 \end{abstract}
 \pacs{46.15.-x,  61.82.Rx, 62.25.+g, 62.30.+d, 63.22.+m} 
 \maketitle
 The deformation of a solid triggers processes which operate across
 several length and time scales. On long length and time scales its
 behavior can be described by a set of hydrodynamic equations
 \cite{Enz74,fleming1976}, which describe, e.g., elastic
 deformation dynamics of the body. On atomic length ($\sim
 10^{-10}$m) and time ($\sim 10^{-13}$s) scales, on the other hand,
 the dynamics can be captured by direct molecular dynamics (MD)
 simulations, which incorporate local bonding information either
 through direct quantum-mechanical calculations or
 semi-empirical many-body potentials. While innovations in
 computing methods have greatly improved the efficiency of MD
 simulations, standard atomistic computer simulations are still
 limited to fairly small system sizes ($\sim 10^9$ atoms) and short
 times ($\sim 10^{-8}$s). This limitation is most severe when
 developing simulation models to study the physics and mechanics of
 nanostructured materials, where the relevant length scales are
 atomic and time scales are mesoscopic. In this regime, the
 available numerical tools are rare.
 
 Progress towards alleviating this limitation has recently been
 made by the introduction of a new modeling paradigm known as the
 {\em phase-field crystal} (PFC) method \cite{elder2002}. This
 method introduces a local atomic mass density field $\rho({\bf
 {r}})$ in which atomic vibrations have been integrated out up to
 diffusive time scales. Dissipative dynamics are then constructed
 to govern the temporal evolution of $\rho$. Unfortunately, the
 original PFC model evolves mass density only on diffusive time
 scales. In particular, it does not contain a mechanism for
 simulating elastic interactions, an important aspect for studying,
 for example, the deformation properties of nanocrystalline solids.

 In this Letter, we introduce a {\it modified phase-field crystal}
 (MPFC) model that includes both diffusive dynamics and elastic
 interactions. This is achieved by exploiting the separation of
 time scales that exists between diffusive and elastic relaxation
 processes in solids. In particular, the MPFC model is constructed
 to transmit long wavelength density fluctuations with wave modes
 that propagate up to a time scale $t_w$, after which the
 strain-relaxed density field continues to evolve according to
 diffusive dynamics. The key feature of our approach is that the
 value of $t_w$ can be chosen to be much smaller than the
 characteristic time scale of diffusion and still much larger than
 $1/\omega_D \sim 10^{-13}$s, where $\omega_D$ denotes the Debye
 frequency.

 The phase field crystal methodology begins by introducing an
 effective free-energy expanded to lowest order in the mass density 
 $\rho({\bf {r}})$:
 \begin{equation}
 F[\rho;T]= \int (\rho/2 [r+(q_o+\nabla^2)^2]\rho +\rho^4/4) d^2x,
 \label{free_energy}
 \end{equation}
 Here, $r \sim (T-T_m)/Lc_P$ and $T_m$, $L$ and $c_P$ are,
 respectively, the melting temperature, the latent heat of fusion
 and specific heat at constant pressure of the pure material. Also,
 $q_o=2 \pi /a$, where $a$ is the equilibrium lattice spacing. This 
 free energy is identical to the one used in Ref.~\cite{elder2002}, 
 and gives rise to a phase diagram of coexisting liquid, solid, and 
 striped phases, as shown in Fig.~1(a). In the solid phase, $\rho$ 
 is non-zero everywhere and spatially periodic on the atomic scale 
 with hexagonal symmetry in two spatial dimensions. In the liquid phase, 
 $\rho$ takes on a constant value everywhere. In the PFC formalism 
 lattice sites are always occupied and vacancy diffusion and topological 
 defects are represented via modulations of the local density amplitude and
 wavelength. Elastic constants can be determined by computing
 $C_{ijkl}=\delta^2 F/\delta u_{ij} \delta u_{kl}$, where $u_{ij}$
 represents the strain of a particular deformation state.

 In the original PFC model, the evolution of the mass density is given by
 \begin{equation}\label{old_pfc}
 {\partial \rho }/{\partial t} =\alpha^2 \nabla^2 \left( {\delta F[\rho;T]
 }/{\delta \rho} \right).
 \end{equation}
 where $\alpha$ is a constant. A severe limitation of the PFC model
 in Eq.(\ref{old_pfc}) is that it only allows for diffusive density
 relaxation. The model does not inherently contain a suitable
 separation of times scales between phase transformation kinetics
 and the much more rapid (``instantaneous'') elastic relaxation.
 This precludes the study of phase transformation phenomena in the
 presence of complex mechanical deformations. It should be pointed
 out that while homogeneous deformations can be imposed through an
 affine transformation \cite{berry2005}, this method is inapplicable 
 in cases where non-homogeneous stress distributions arise.
\begin{figure}[t]
 \includegraphics*[width=3in,height=3in]{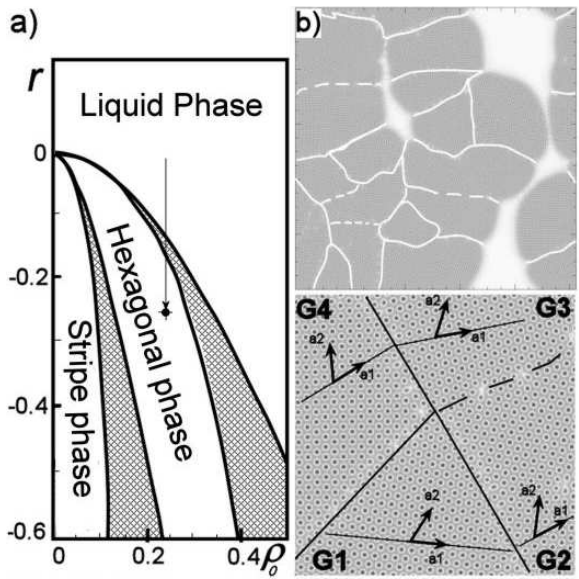}
 \caption{\label{gb} (a) Phase diagram indicating isothermal quench.
 The diagram is symmetric around $\rho_0=0$. (b-top) Snapshot in the
 evolution of polycrystalline solidification using the MPFC model.
 Grain boundaries are highlighted in white. (b-bottom) Zoom-in of 4
 crystal grains and their orientations. }
 \end{figure}
 As will be demonstrated below, these serious shortcomings of the
 original PFC model can be circumvented in a way that allows us to
 preserve the quintessential advantage of the PFC model, namely, the
 ability to simulate atomic-scale interactions and dynamics on time
 scales many orders of magnitude longer than molecular dynamics
 time scales. Most importantly, our modified model naturally
 incorporates ``instantaneous'' elastic interactions.

 We begin by introducing a modified PFC equation given by
 \begin{eqnarray}
 \frac{\partial^2 \rho}{\partial t^2} +\beta \frac{\partial
 \rho}{\partial t}= \alpha^2 \nabla^2 \mu + \xi \label{MPFC_main}
 \end{eqnarray}
 where $\mu=\delta F[\rho;T]/\delta \rho$, while $\beta$ and
 $\alpha$ are phenomenological constants. $F$ is the free-energy
 introduced in Eq.~\ref{free_energy} and $\xi$ is a Gaussian random
 variable with correlations satisfying
 $\langle \xi \xi' \rangle= k_B T \beta q_o^{d-4}/ \lambda^2
 \nabla^2 \delta(\vec{x}-\vec{x^{\prime}}) \delta(t-t^{\prime})$.
 Henceforth, we will set $\xi=0$. Equation~\ref{MPFC_main} is of
 the form of a damped wave equation, containing two propagating
 density modes at early time and one diffusive mode at late times.
 Specifically, the fast dynamics of the MPFC model are governed by
 the first term of Eq.(\ref{MPFC_main}), while the late time
 dynamics are governed by Eq.(\ref{old_pfc}).

 To elucidate the dynamics described by Eq.~ (\ref{MPFC_main}), we
 performed a Floquet stability analysis. This was done by assuming
 a perturbation in the density of the form $\rho_p=\rho_{eq}+\delta
 \rho$, where $\rho_{eq}=\rho_o+\sum_{n,m} a_{n,m}
 e^{i\vec{G}_{n,m}\cdot \vec{r}}$, with $\rho_o$ the average
 density, $\vec{G}_{n,m}=n \hat{x} + (n+2m)/\sqrt{3} \hat{y}$ the
 triangular reciprocal lattice vectors and $a_{n,m}$ their
 corresponding amplitudes. Here, $\delta \rho=\sum_{n,m} b_{n,m}(t)
 e^{ i \vec{G}_{n,m}\cdot\vec{r}+i\vec{Q}\cdot\vec{r}}$, where
 $\vec{Q}$ is a perturbation wave vector and $b_{n,m}(t)$ the
 perturbation amplitude of mode $(m,n)$. Substituting $\rho_p$ into
 the model and expanding to linear order gives an equation for
 $b_{n,m}$. The leading order mode satisfies $b_{0,0} \sim e^{i
 \omega t}$. The dispersion relation $\omega(Q)$ is given by
 $\omega(Q)= i \beta /2 \pm \Lambda(Q)/2$, where $\Lambda(Q) =
 \sqrt{-\beta^2 + 4 \alpha^2 Q^2 \left[ 3\rho_o^2 + r +
 (Q^2-q_o^2)^2 + 9/8 A_{min}^2 \right]}$. Note that when $4
 \alpha^2 Q^2 \left[ 3\rho_o^2 + r + (Q^2-q_o^2)^2 + 9/8 A_{min}^2
 \right] \gg \beta^2$, the dispersion is approximately $\omega(Q)
 \approx i \beta /2 \pm 2 \alpha Q \sqrt{3\rho_o^2 + r +
 (Q^2-q_o^2)^2 + 9/8 A_{min}^2} \equiv i \beta/2 \pm v_{eff} Q$.
 This dispersion describes a pair of waves that propagate undamped
 for time $t_w \approx 2 \beta^{-1}$ and distance $L \sim v_{eff}
 t_w$, after which they become effectively diffusive as in
 Ref.~\cite{elder2002}. It is precisely these propagating modes
 which mediate elastic interactions in the model. Details of this
 calculation as will be presented elsewhere ~\cite{papertoappear}.

 This analysis demonstrates that Eq.(\ref{MPFC_main}) allows us to
 transmit disturbances across long distances using wave modes that
 propagate on a time scale $t_w$. This allows all atomic positions
 to relax to a position close to their deformed equilibrium
 positions prior to any significant diffusion taking place. Most
 notably, the time scale $t_w$ can be set many orders of magnitude
 larger than characteristic vibrational time scales ($\sim
 10^{-13}$s) but still significantly faster than the scale on which
 diffusive processes occur.

 We now turn to the treatment of the fully non-linear evolution 
 of Eq.~(\ref{MPFC_main}). The details of our numerical procedures are 
 as follows. All simulations were conducted on a rectangular grid using
 periodic boundary conditions. Space was measured in units of the
 lattice constant $a$, while the grid size $\Delta x$, time step
 $\Delta t$ and coefficients $\alpha$, $\beta$ were chosen
 according to the particular application. External loads were
 applied to the boundary of our system by using a {\it penalty
 function method}. In this method, an additional term, of the form
 $P=M(x,y,t)\sqrt{(\rho-\rho_{bdy})^2}$, is added to the free
 energy. This term couples the sample density $\rho$ to an imposed
 periodic density field $\rho_{\rm bdy}$. The support of $\rho_{\rm
 bdy}$ is the same as the support of the function $M(x,y,t)>0$,
 which defines the shape of the desired loading surface. The form
 of $P$ thus couples some portion of the sample's density (e.g.
 near the sample boundaries) to the imposed boundary potential
 $\rho_{bdy}$, which results in the sample's density field
 becoming slaved to the peaks of $\rho_{\rm bdy}$ as $|M(x,y,t)|
 \rightarrow \infty$.  As the applied potential field is
 translated, the sample's density field, along the loading surfaces,
 adiabatically follows the applied field. This specific form of $P$
 also assures that our penalty function does not alter the
 equilibrium phase diagram of the basic free energy functional
 $F[\rho;T]$ defined above.

 We first simulated isothermal solidification using
 Eq.(\ref{MPFC_main}) by preparing the system in the liquid state
 and subsequently setting the temperature below the coexistence
 line in the phase diagram. In this case, the dynamics gives rise
 to nucleation and growth of the solid phase in the presence of
 thermal fluctuations. To facilitate nucleation, several nucleation
 sites were initiated in the metastable liquid phase in the form of
 random (Gaussian) fluctuations. During solidification, we found
 that the effect of the first term in Eq.(\ref{MPFC_main}) was
 negligible, and the growth rates and morphology were essentially
 indistinguishable from those using Eq.(\ref{old_pfc}).
 Figure~\ref{gb} illustrates growth and impingement of several
 nuclei in an undercooled melt. The simulation was started with the
 liquid of average density ${\rho}_{0}=0.285$ and dimensionless
 temperature $r=-0.25$; other parameters were set to $(\Delta x,
 \Delta{t}, \alpha, \beta)=(\pi/8, 0.001,15,0.9)$. The measured grain boundary
 energies per unit length are consistent with the usual
 Read-Shockley form \cite{elder2003,War03}

\begin{figure}[t]
 \includegraphics*[width=3.5in]{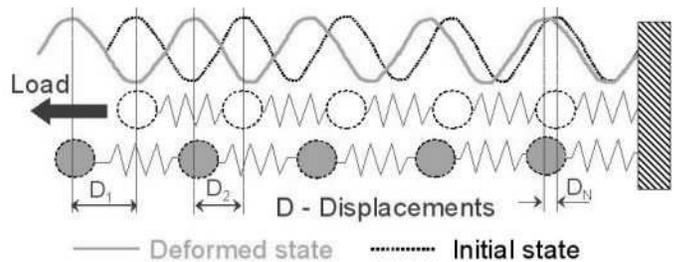} \caption{\label{spring}
 Schematic illustration of the atomic locations before and after
 tension is applied in the MPFC model.}
 \end{figure}

\begin{figure}[t]
 \includegraphics*[width=3.5in]{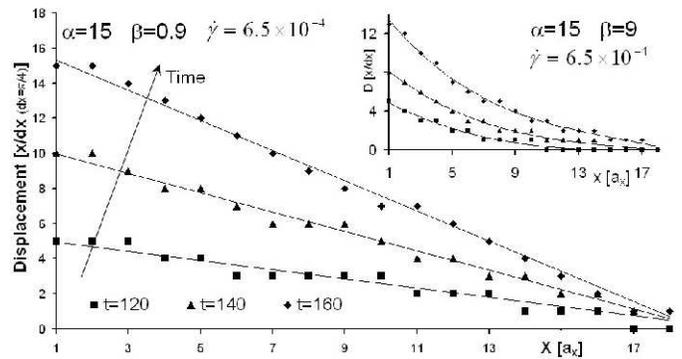} \caption{\label{displacements}
 The displacements along a one-dimensional sample in simple
 uniaxial tension at three different times (top). Linear profiles
 are consistent with linear elasticity theory. Inset: a ten-fold
 increase in $\beta$ leads to visco-elastic behavior.}
 \end{figure}

 To demonstrate the presence of elastic relaxation modes in the
 MPFC model, we performed simulations of an effectively one
 dimensional single-crystal specimen under uniaxial tension. The
 system was prepared in the coexistence region as given by the
 phase diagram, and the solid sample was surrounded by liquid.
 Model parameters used were $(r, \psi, \Delta x, \Delta{t},
 \alpha, \beta)=(-0.4, 0.31, \pi/8, 0.001,15,0.9)$. A tensile load
 applied to a semi-infinite continuum elastic bar can be
 theoretically modelled as array of coupled masses and springs
 along the $x$-axis, as illustrated in Fig.~\ref{spring}. When an
 atom at the boundary is displaced by an amount $D_1$ to the left,
 a tensile stress wave will propagate to the right. When atomic
 oscillations stop, a linear displacement distribution, $D(x)=D_1 x/L$,
 will be established along the bar. Plots of displacement
 vs.~ position in the case of constant strain rate applied to the
 boundary atom are shown in Fig.~\ref{displacements} at three
 different times. Here, the displacements were extracted by a peak
 tracking method, where the locations of local maxima in $\rho$
 were tabulated after each time step. The data clearly shows that
 the response of the system is consistent with elasticity theory.
 \begin{figure}[b]
 \includegraphics*[width=3.5in]{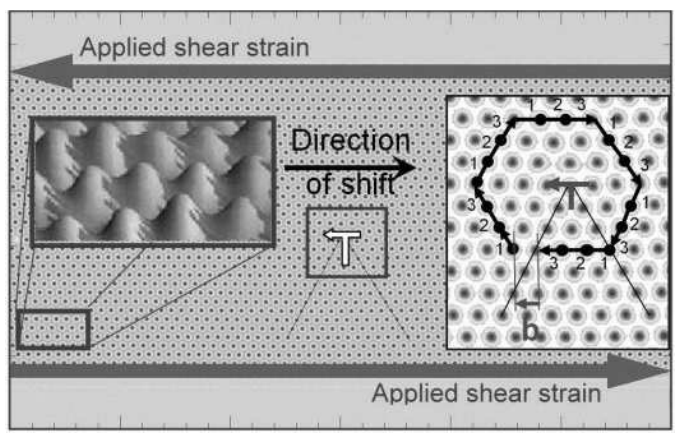}
 \caption{\label{dislocation} A portion of the sample used to examine
 dislocation glide velocity. Parameters used: $(r, \psi, \Delta{x},
 \Delta{t}, \alpha, \beta)=(-1, 0.49, \pi/4, 0.001, 15 ,0.9)$.}
 \end{figure}
 \begin{figure}
 \includegraphics{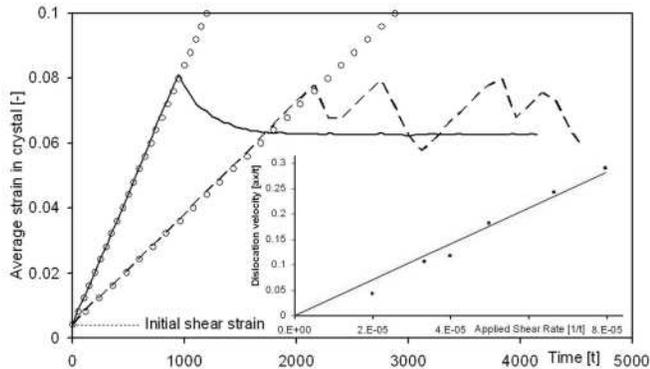}
 \caption{\label{disl_strain}Two regimes of dislocation glide. For
 high strain rates we observed continuous glide, while at lower
 strain rate the dislocation set into a stick-slip motion. Inset:
 Dislocation glide velocity Vs. applied strain rate. }
 \end{figure}
\begin{figure}
 \includegraphics[width=3in]{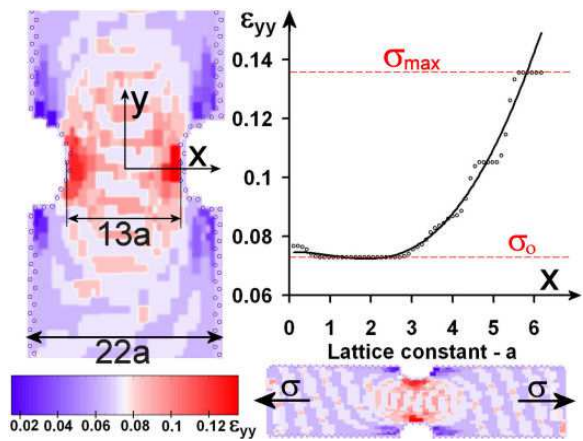}
 \caption {\label{stress_conc} Strain concentration in a double
 notched sample under a uniaxial tension. \textbf{Left:} A strain
 map of the center portion of the sample displayed at the bottom. Boundary atoms are
 highlighted in black. \textbf{Right:} Plot represents a strain
 profile from the center of the sample into the root of the notch.
The solid line is a guide to the eye.}
 \end{figure}
 To make contact with the previous PFC formulation in
 Eq.(\ref{old_pfc}) \cite{elder2003}, we repeated the same
 simulations with a ten-fold increase in the damping parameter
 $\beta=9$. The computed displacements, plotted in the inset of
 Fig.~\ref{displacements}, show that the response becomes
 viscoelastic as damping is increased. Therefore,
 Eq.(\ref{old_pfc}) alone does not adequately describe elastic
 responses in strained crystals at finite strain rates, while
 Eq.(\ref{MPFC_main}) naturally incorporates such phenomena. We
 note that for simple modes of deformation, Eq.(\ref{old_pfc}) can
 still be used to model elastic relaxation by uniformly adjusting
 all the atomic peaks of the density field $\rho$ after a specified
 number of numerical time steps, which is equivalent to carrying
 out an affine transformation. This method is applicable in, e.g.,
 elucidating the glide dynamics of a single dislocation
 \cite{berry2005}. However, this approach cannot be used to handle
 elastic relaxation in the case of complex geometries, non-uniform
 stresses, and high strain rates \cite{Sch03,Van99,Yam02,Rem04}.

 We also examined the dynamics of individual dislocations. The
 set-up for these simulations are shown in Fig. \ref{dislocation}.
 Specifically, the top part of the crystal initially contains N
 atoms and the bottom part N+1. After the sample equilibrated an
 edge dislocation formed and a constant shear strain rate was
 applied. The time-averaged dislocation glide velocity $\bar{v}$
 was found to be a linear function of the strain rate $\dot{\gamma}$, 
 consistent with classical dislocation theory. This theory predicts 
 that $\bar{v}=\dot{\gamma}/(\rho_d b)$, where $\rho_d$ is the dislocation 
 density and $b$ is the magnitude of the Burger's vector~\cite{johnstongilman59}.

 In order to elucidate the local dynamics of individual
 dislocations, we computed the average strain in the crystal as a
 function of time for different strain rates. These results, shown
 in Fig.~\ref{disl_strain}, revealed two regimes of dislocation
 glide. The first was characterized by continuous glide (observed
 at large $\dot{\gamma}$) and the second by a stick-slip gliding of
 the dislocation at low $\dot{\gamma}$. In both cases the applied
 plastic strain was relieved by the motion of the dislocation, and
 the time-averaged strain remained constant.

 To further illustrate the properties of our MPFC model, the effect
 of uniaxial tension in a notched sample was examined.
 Figure~\ref{stress_conc} shows that strain (stress) in a notched
 sample appropriately concentrates near the notches, as expected
 from linear elasticity theory. In particular, treating the case of
 a double notched plate the stress concentration for this geometry
 is $K_t=\sigma_{yy}^{max}/\sigma_{yy}^{0} = 1.8$ \cite{peterson53}, 
 which is in excellent agreement with our simulation result $1.81$ . It is 
 noteworthy that a simulation with the PFC model (Eq.~\ref{old_pfc})
 for the same system and using an affine transformation to approximate 
 the strains in the sample, failed to produce the expected strain 
 concentration.

 In conclusion, we have introduced a novel phase-field crystal
 model (MPFC), which extends the previous phase-field crystal
 formalism by generating dynamics on two time scales. Atomic
 positions are relaxed rapidly at early times in a manner
 consistent with elasticity theory, while late time dynamics are
 governed by diffusive dynamics characteristic of phase
 transformation kinetics, vacancy diffusion, grain boundary
 kinetics and dislocation climb. It is expected that the MPFC model
 will help open a new window into the study of phase transformation
 kinetics and microstructure heterogeneity in high strain rate
 loading of nanocrystalline solids.
 
 \begin{acknowledgments}
 This work has been in part supported by the National Science and
 Engineering Research Council of Canada (NP) and an NSF-DMR Grant
 No.~0449184 (MH).
 \end{acknowledgments}

 \end{document}